\documentclass[useAMS,usenatbib,usegraphicx,fleqn,a4paper]{mn2e}
\usepackage{times}

\pdfpagewidth=\paperwidth
\pdfpageheight=\paperheight

\usepackage{graphicx}
\usepackage{amsmath}
\usepackage{subcaption}
\usepackage{import}
\usepackage{mathptmx}
\usepackage{inputenc}
\inputencoding{utf8}

\usepackage[x11names, rgb, svgnames]{xcolor}
\usepackage[colorlinks=True,
citecolor=blue,
linkcolor=blue,
pdfauthor={Janis Hagelberg},
pdftitle={The Geneva Reduction and Analysis Pipeline for High-contrast Imaging of planetary Companions},]{hyperref} 

\usepackage{setspace}

\title[GRAPHIC]
   {The Geneva Reduction and Analysis Pipeline
for
High-contrast Imaging of planetary Companions}

\author[J. Hagelberg,
       D. S\'egransan,
       S. Udry,
       and F. Wildi]
{J. Hagelberg$^{1,2}$\thanks{E-mail: janis.hagelberg@unige.ch},
       D. S\'egransan$^{2}$,
       S. Udry$^{2}$,
       and F. Wildi$^{2}$\\
$^{1}$Institute for Astronomy, University of Hawai'i, 2680 Woodlawn Drive,
Honolulu, HI 96822, USA\\
$^{2}$Observatoire de Gen\`eve, Universit\'e de Gen\`eve,
     51 Chemin des Maillettes, 1290, Versoix, Switzerland}

\begin{document}

\date{Accepted 2015 October 14.}

\pagerange{\pageref{firstpage}--\pageref{lastpage}} \pubyear{2015}

\maketitle

\label{firstpage}

\begin{abstract}
We present  {\small GRAPHIC}, a new angular differential imaging (ADI)reduction pipeline where all geometric image operations are based on Fourier transforms.
To achieve this goal the entire pipeline is parallelised making it possible to reduce large amounts of observation data
without the need to bin the data. The specific rotation and shift algorithms based on Fourier transforms are described
and performance comparison with conventional interpolation algorithm is given. Tests using fake companions
injected in real science frames demonstrate the significant gain obtained by using geometric operations based on Fourier transforms
compared to conventional interpolation. This also translates in a better point spread function and speckle subtraction
with respect to conventional reduction pipelines, achieving detection limits comparable to current best performing pipelines.
Flux conservation of the companions is also demonstrated.
This pipeline is currently able to reduce science data produced by VLT/NACO, Gemini/NICI, VLT/SPHERE, and Subaru/SCExAO.
\end{abstract}


\section{Introduction}

Eighteen years after the first discovery of an exoplanet around a sun-like star \citep{mayor_jupiter-mass_1995} and
the unambiguous detection of three brown dwarfs \citep{basri_surprise_1995,
  nakajima_discovery_1995,rebolo_discovery_1995}, thousands of planets and brown dwarfs have been discovered. These numbers are growing ever
faster as the pace of new detections is increasing, thanks to newly built instruments purposely designed to search for
sub-stellar objects, but also to the optimisation of data analysis techniques.

The vast majority of exoplanets are currently detected with the radial velocity or transit techniques.
However, orbital periods longer than the time-span of the observations will hardly be
detected by these two techniques, inducing a sharp decrease in detectability beyond $\approx\,5$ AU and leaving unprobed a large area in
the mass-separation parameter space.
Direct imaging on the opposite probes the outer orbital regions not accessible with the two previous techniques, but the
high contrast at small separation which needs to be reached turns it into one of the most challenging exoplanet
detection techniques. The main hurdle to detect companions by high contrast imaging is to remove the stellar point
spread function (PSF)
without diminishing the signal from the faint companion.
This can be achieved through instrumental improvements
or by improving the observing and data reduction techniques, with efforts focusing on these two fronts concurrently.

Since the first planets around stars have been directly imaged \citep{marois_direct_2008, lagrange_probable_2009}, the
rate of exoplanets detected by direct imaging is steadily increasing
thanks to the progress made to overcome the many technical challenges and careful selection of the target samples.
But the small number of detections in total contrasts with the many direct imaging surveys which generated only few or no detection at all (e.g., \citealt{ masciadri_search_2005,
  biller_imaging_2007, lafreniere_gemini_2007, chauvin_deep_2010, heinze_constraints_2010, vigan_international_2012,bowler_planets_2012,
  nielsen_gemini_2013, wahhaj_gemini_2013, janson_seeds_2013, crepp_trends_2012}).

The technical challenge of subtracting the host star point spread function is currently based on two complementary differential imaging
methods, with the
same core idea of generating a point spread function as similar as possible to the one which should be subtracted, but without having any potential
companion signal in it. The difficulty being that the speckle structure of the point spread function evolves in time, with many speckles
in the stellar halo having a similar shape and intensity to what would be expected from a companion.
Two main differential techniques exist to detect thermal emission of a companion.
The first method, called Simultaneous Differential Imaging (SDI) is based on simultaneous observations in multiple bands. One can then either take
advantage of specific absorption bands of the companion so that it is visible in one band and not the other,
thus making it possible to subtract the PSF while keeping the companion signal, or use the fact that because speckles are chromatic, their pattern scales with the wavelength but the
potential companion stays on the same spot  \citep{racine_speckle_1999, lenzen_novel_2004}. The other differential method,
known as Angular Differential Imaging (ADI) is based on the rotation of the field
\citep{schneider_coronagraphy_2003, liu_substructure_2004, marois_angular_2006}, and has proven to be currently the most efficient method for point spread function
subtraction when searching for companions. These two methods do not require the use of a coronograph even though their use can increase the detection
limits in certain cases.
Finally, the two methods can be combined by letting the field rotate while observing simultaneously in multiple bands.
Nearly every survey developed its own reduction
pipeline most often based on either the Locally Optimized Combination of Images (LOCI) \citep{lafreniere_new_2007}
or more recently Principal Component Analysis (PCA) \citep{soummer_detection_2012, amara_pynpoint:_2012}.

Here we present the Geneva Reduction and Analysis Pipeline for
High-contrast Imaging of planetary Companions based on ADI for point spread function subtraction, which makes intensive use of Fourier
analysis. It was specifically developed for the Geneva high-contrast imaging search of
companions revealed by radial velocity trends in the HARPS and CORALIE survey.

\section{The Geneva high-contrast imaging search of companions revealed
by radial velocity trends in the HARPS and CORALIE surveys \label{sec:survey}}
Our campaign aims at detecting with direct imaging companions revealed by the RV trend they are
causing, based on data from our two CORALIE and HARPS RV planet-search surveys. The radial velocity data spans over more than a decade
with a precision reaching below 1 m/s in the case of HARPS so that trends induced by sub-stellar companions on wide
orbits can readily be
detected. The selected targets are observed using VLT/NACO and the angular differential imaging technique with deep
observations of up to four hours on target in order to reach the faint companions which had time to cool down.
Our targets are all bright which results in integration times below one second to reach saturation, and in order not to resort to
frame binning we are using the cube mode offered by NACO where frames are stacked into a data cube. Each cube containing
hundreds of frames is then saved into a single {\small FITS} file with the benefit of reducing readout overheads during observations.

\subsection{Parallelisation}

The NACO observation sequences of up to four hours used in our campaign lead to roughly 100GB of data and 100'000 frames. A straightforward single core
reduction would take an extremely long time and would run out of memory before finishing, due to the many complex
 operations involved in the data reduction mostly based on Fast Fourier Transforms ({\small FFT}).
The most widely used and easiest solution to this large data handling issue would be to average bin the data.

By suitably binning the frames, one can decrease the total amount of data to a quantity which fits the hardware limitations.
The drawback is that valuable information gets lost in the binning process. First of all, the characteristics of atmospheric turbulence are not constant in time
as well as the quality of the adaptive optics (AO) turbulence correction. The Strehl ratio for binned frames is the mean ratio of the frames in
the bin, so that if half of the frames have poor adaptive optics correction the final binned frame will also have below average Strehl,
even though the other half of the frames had good Strehl. Furthermore, binning frames before re-centring and correcting for the field rotation
smears out the companion point spread function which in turn decreases its signal in the final product. This is why we decided not to
rely on binning.

\begin{figure*}
\subcaptionbox{Spatial parallelisation\label{fig:data_para_spatial}}[0.3\textwidth]
{\includegraphics[width=0.3\textwidth]{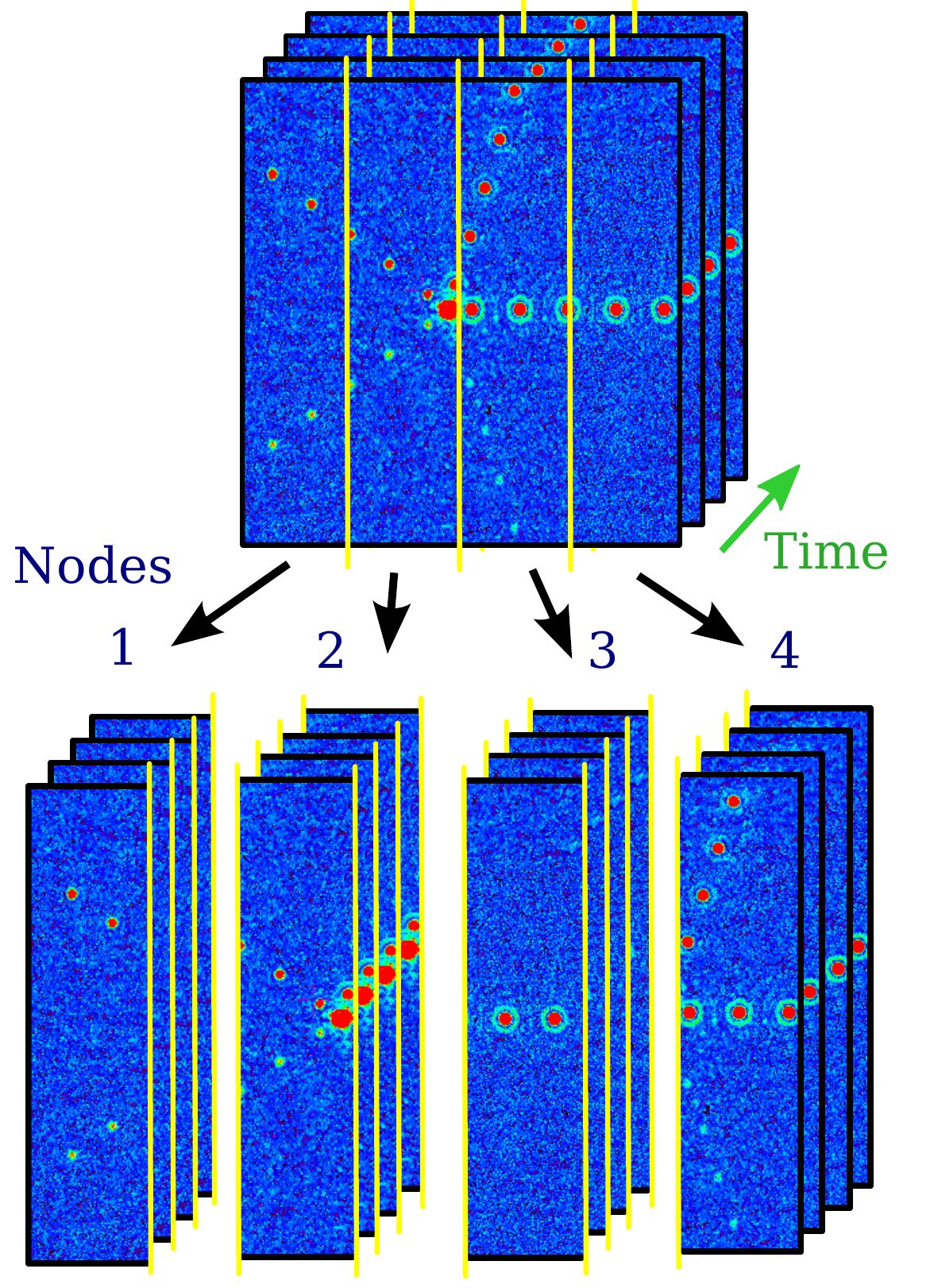}
}
\subcaptionbox{Time parallelisation\label{fig:data_para_temporal}}[0.5\textwidth]{
\includegraphics[width=0.5\textwidth]{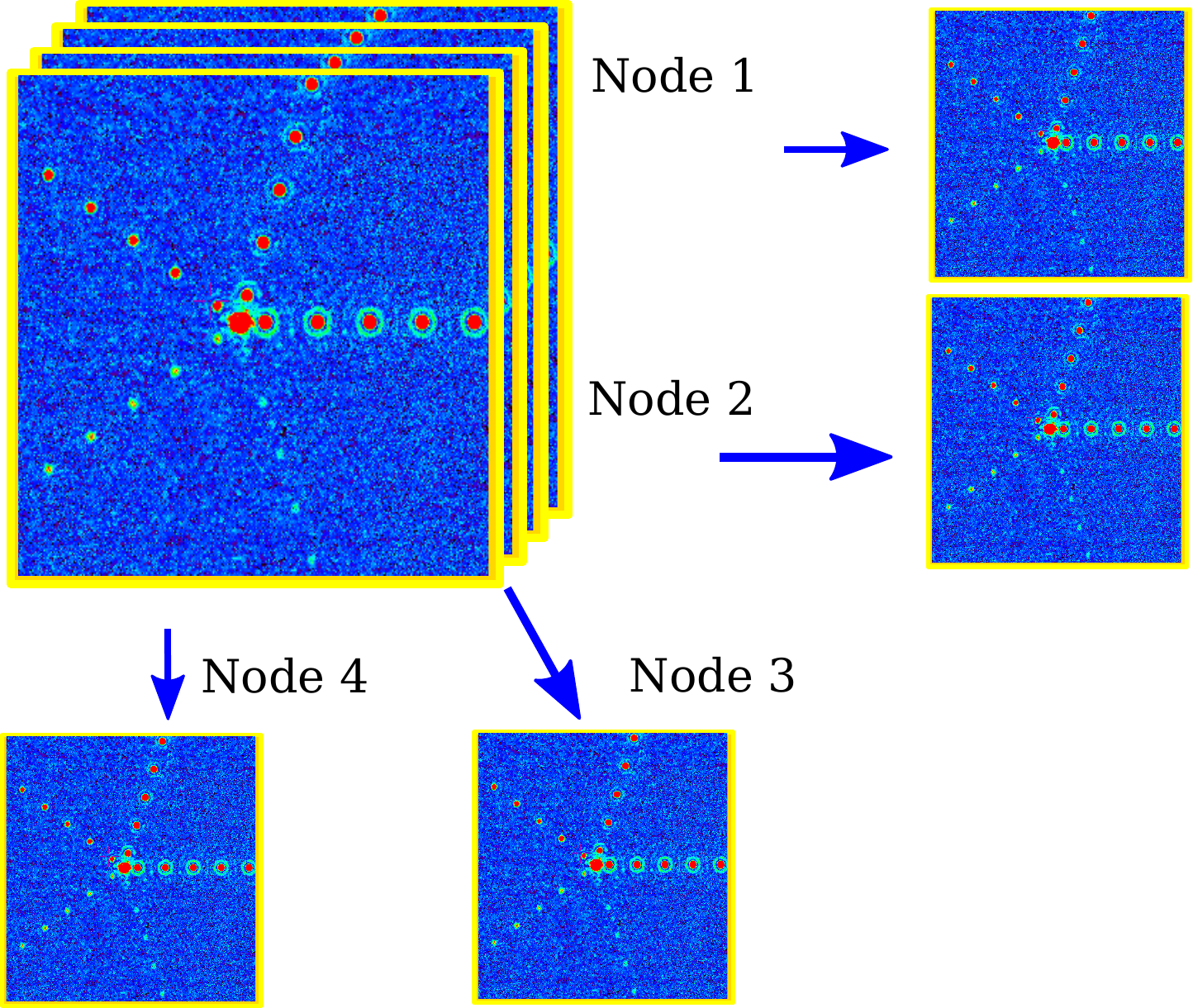}%
}
\caption{The two different data parallelism schemes used by the pipeline. In spatial parallelism (a), the observation frames are
  cut along the time axis. Each node is then processing a different region of the frame. In time parallelisation (b), the frames are
  distributed among the nodes. In that case each node processes a different frame, or rather frame group.}\label{fig:data_paral}
\end{figure*}

The different algorithms of our pipeline fit very well to a \emph{data parallelism} scheme, which focuses on distributing the data
across different parallel computing nodes. A master node shares the data between the slave nodes, which
only have a fraction of the data to process. Once the slaves have finished, the data are gathered by the master and reassembled.
Two different types of parallelisation are used, which differ on the way the data are shared between the nodes. If the
operations are pixel based, the spatial parallelisation scheme is used. In this scheme the data cube is cut in pieces
along the time axis, which means that each node receives one specific region of all the frames (see
\autoref{fig:data_para_spatial}).
The other scheme, temporal parallelisation, is used when the whole frame is needed for a specific operation. This is for
example the case for shifts and rotations. The data cube is then separated in frame packages, and each node receives a
different package containing full frames (see \autoref{fig:data_para_temporal}).

The pipeline is implemented in {\small PYTHON} using {\small C} and {\small FORTRAN} libraries for the calculation intensive parts.
The parallelisation is achieved using the \emph{Open Message Passing Interface ( {\small OPENMPI}}, \citealt{gabriel_open_2004}). Parallelisation can be distributed
transparently
among many different nodes, independently of their architecture. The interface between the {\small PYTHON} code
and {\small OPENMPI} is handled by the {\small MPI4PY} module \citep{dalcin_mpi_2008}.
All the data reduction steps given in this paper are parallelised, and based on the specificity of the process either in spatial or time parallelisation.

\section{ADI data reduction}

The pipeline was initially developed to reduce non-coronographic saturated ADI observations in \textit{L'} band ($3.8\mu m$)
 from the CONICA 1024x1024 InSb Aladdin 3 detector which is part of the NACO instrument at the VLT \citep{lenzen_naos-conica_2003}.
It was later extended to process coronographic data from NACO,
GEMINI/NICI \citep{chun_performance_2008}, VLT/SPHERE
\citep{beuzit_direct_2010} and Subaru/SCExAO \citep{jovanovic_subaru_2015} in any band.

The concept of this pipeline is to preserve the companion photometry by relying on techniques which have the least
impact on the signal. Conservation of the noise structure is also a priority in order to carry out an efficient noise subtraction.
This is obtained by applying the geometric transformations for centring and rotation in Fourier space. Another important
aspect of the pipeline is the scalable parallelisation, which makes it possible to run it on anything
from a cluster for high performance computing (HPC) down to any modern laptop.

The reduction procedure is partitioned in four main parts, which will be described hereafter:
\begin{enumerate}
\item registration \ref{sec:registration}
\item image pre-processing \ref{sec:image_pre_processing}
\item PSF subtraction \ref{sec:psf_sub}
\item derotation \ref{sec:derotation}
\end{enumerate}

\subsection{Registration}\label{sec:registration}

The very first step of the reduction process is the registration. Every data frame is analysed
and a table is generated containing for every frame the time of exposure, the parallactic angle, the star centre and other
point spread function characteristics needed for the different reduction steps.
Even though the data cubes are not modified at this step, it is a key element of the reduction as any error on the angle
or star centre determination smears out the companion signal.

The registration process is also used to assess the quality of the adaptive optics correction for every single frame.
This quality estimate is then later used to keep only the best frames by manually setting the constraints on the frame quality selection.
This results in a kind of lucky adaptive optics.

\subsubsection{Parallactic angle}

The parallactic angle varies in time as the star moves in the sky, and its hour angle changes. Based on the
``Local Sidereal Time'' (LST) given in seconds in the FITS header, and the right ascension $\alpha$ of the star
(given in degrees), the hour angle $h$ is given by:

\begin{center}
\[
\centering
h= \frac{15\cdot {\mbox{LST}}}{3600}-\alpha.
\]
\end{center}

When observing with NACO in cube mode the frames are stored in data cubes of 30 to more than 500 frames, with only one
single header. This implies that the hour angle for each single frame cannot directly be derived
from the header, since solely one LST value is given.
Thus the observing time for every single frame has to be interpolated in order to get the correct parallactic angles, which
can become problematic when frame loss occurs.

\subsubsection{Star point spread function registration}
The determination of the star centre position is a complex task
particularly in the case of saturated point spread function and/or
coronographic imaging because the information at the core of the
point spread function is often lost. The determination of the cen-
tring method accuracy is also limited by the fact that the exact
point spread function centre is unknown. In order to make the
point spread function registration work on any combination of sat-
urated/unsaturated, full/coronographic, and AO/non-AO modes the
algorithm is split in a two stage process. First a basic centroid
search is done, and once the centroid is found a two-dimensional
function is fitted to the point spread function, which takes into ac-
count possible coronographic or saturated cores by masking out
pixels. This method also works if there is more than one star in the
field of view, as long as no other star has exactly the same flux as
the target.

The first step which is the centroiding algorithm searches for
a patch of contiguous pixels above a given threshold value. If the
patch size is within the range given by the user, the centre-of-mass
of the patch is calculated. These values are then fed into the point
spread function fitting algorithm as initial values.

The pipeline uses the following form of the \citet{moffat_theoretical_1969} function to fit the
point spread function centred on $x_0,y_0$:

\[
I(x,y,\alpha, \beta )=I_0\cdot\frac{\beta -1}{\pi
  \alpha^2}\cdot\left(1+\frac{(x_0-x)^2+(y_0-y)^2}{\alpha^2}\right)^{-\beta}+{\mbox{\small Bg}},
\]

with $\alpha$ the seeing radius parameter, $\beta$ the wing shape parameter, and {\mbox{\small Bg}} the value of the background flux \citep{trujillo_effects_2001}.
The full width at half-maximum (FWHM) of the point spread function is then given by the two fitting variables $\alpha$ and $\beta$:
\[
{\mbox{FWHM}}(\alpha, \beta )=2\alpha\sqrt{2^{1/\beta}- 1}.
\]

The Moffat fitting on the point spread function is achieved using lmdif's modified Levenberg-Marquardt algorithm which is part of the Fortran MINPACK library \citep{more_user_1980}.
The registration procedure achieves an accuracy better than 0.5 pixels on the star centre position, but its efficiency is highly dependent on the observational setup.

With the growing tendency to highly saturate the point spread function core or to use new generation coronographs, it becomes increasingly difficult to fit a Moffat function. A solution
to this is the usage of satellite speckles pioneered by \citet{sivaramakrishnan_astrometry_2006} using a reticulated wire grid, and recently brought to a new state of sophistication by
 \citet{jovanovic_artificial_2015} using
incoherent speckles generated with a pattern on the adaptive optics deformable mirror. Due to the chromaticity of speckles the satellites appear elongated on broad band imaging giving them
a shape which is difficult to fit with a conventional point spread function model. On the other hand, the centroiding algorithm is a good candidate to implement a registration
algorithm based on such satellite speckles.

\subsubsection{Frame selection}

In order to increase the signal to noise ratio of the companion, we keep only the frames with good adaptive optics correction. A
first rough selection is based on the centroiding algorithm, by simply discarding frames where no centroid has been found.

The second step is based on the point spread function geometry. When the detector integration time is more than a few seconds it can
happen that bad tip-tilt correction create a sharp elongated point spread function specially when in coronographic mode. Simple selection on
maximum signal strength hardly detects this kind of frames, even though
the point spread function shape is asymmetric. Using the values of the point spread function fitting has shown to be a robust way to verify the quality of
the adaptive optics correction. Alternatively the satellite speckles position and intensity can be used when available.

Furthermore, by using individual frame point spread function fitting instead of the widely used cross-correlation method, we can ascertain that only
good and optimally centred frames are used. The independent centring also ensures that poor centring on one frame will
not affect centring on the other frames.

Once the frame selection has been done an optional quick-look algorithm can be run which bins the re-centred good frames
to decrease the total number of frames to process in the following steps.

\subsection{Image quality pre-processing}\label{sec:image_pre_processing}

\begin{figure}
\begin{center}
 \includegraphics[width=0.48\textwidth]{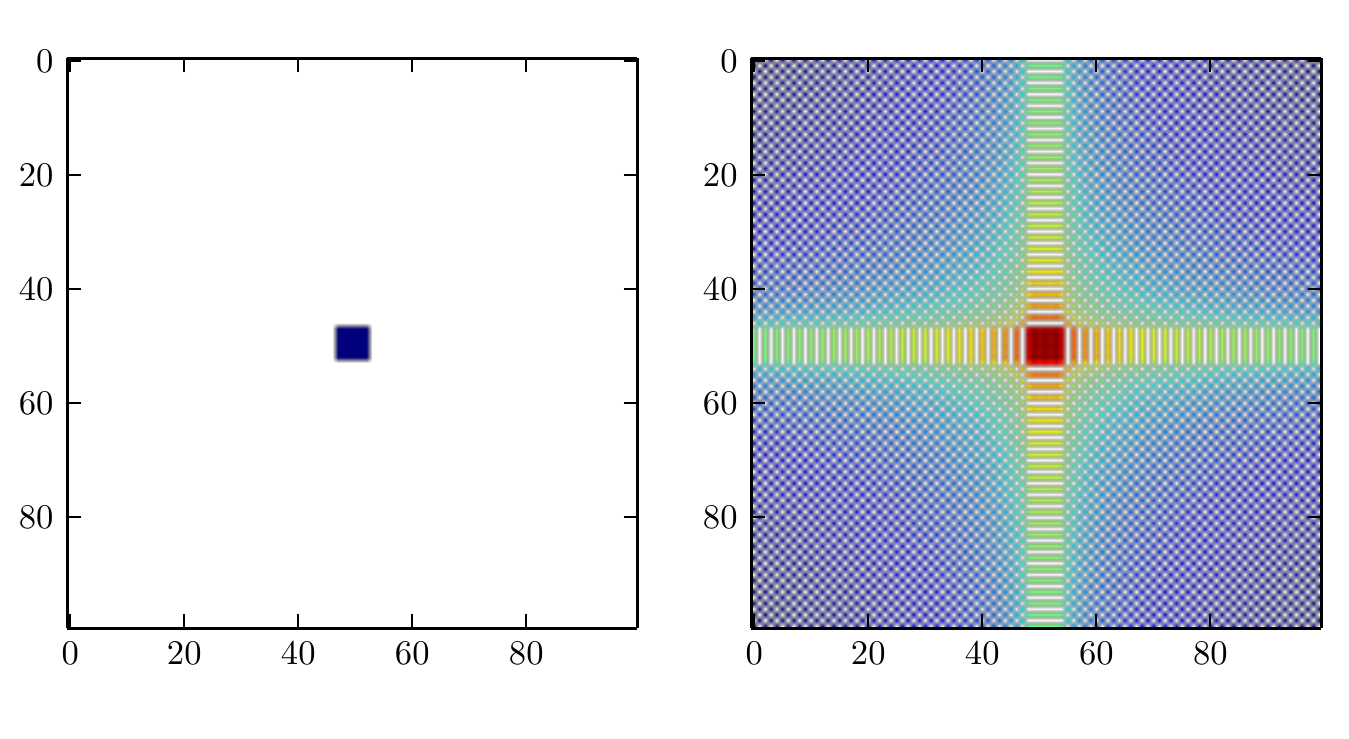}
\caption{The single non-zero pixel surrounded by a background at zero (left figure) is shifted in
Fourier space by $\Delta x=0.3$, and $\Delta y=1.5$ pixels (right figure), resulting in the typical pattern of the Gibbs phenomenon.}
 \label{fig:gibbs}
 \end{center}
\end{figure}

Image quality plays a key role when using Fourier transforms, because of the
\citet{gibbs_fouriers_1898} phenomenon. This phenomenon causes a chequerboard pattern to appear with pixels alternately
overshooting in the positive and
negative values around an image discontinuity, such as bad pixels, saturated point spread function or sharp image borders
 (\autoref{fig:gibbs}).
The reason for these oscillations is that a discontinuity would need infinite Fourier series to be correctly characterised, but
since we are dealing with finite discrete Fourier transforms the discontinuity is not well approximated.

The frame preparation is thus a key step in order to use Fourier-based operations, where no deviant pixel should be left
out. The two key steps for the image pre-processing are the sky generation and subtraction and the bad pixel correction.

\subsubsection{Master sky}

The sky background varies very rapidly when observing in the near infra-red especially in \textit{L'} band. Observations are done
in dithering mode in order to directly use the science frames to determine the sky background.
To decrease the effect of the star point spread function on the sky determination, a mask is applied on the region containing the star.
A median of \emph{N} time-contiguous cubes with \emph{N} different masked dithering positions is then calculated.
As dithering positions change every one or two seconds, and we are using four to five dithering positions,
a sky frame is produced for every five to ten seconds of observation. Using this method we manage to have a median sky as close
as possible to the sky background of the science frames.
Once the master skies have been generated, they are subtracted on each frame, using each time the
master sky frame closest in time.

\subsubsection{Bad pixel correction}

First a bad pixel map is generated using either a master dark or a master sky. A simple sigma clipping routine flags the
deviant pixels, where bad pixels are defined as a pixels
which values varies more than $C\cdot\sigma$ from the median value of the frame.
By changing the value of the $C$ coefficient the selection criteria can be adapted to ensure that all bad pixels are cleaned.

Since each bad pixel will be spread over many pixels as a consequence of the sub-pixel re-centring and the derotation
process, it is not possible to simply leave out a bad pixel.
Not to mention the Gibbs phenomenon such pixels would induce.
To clean the bad pixels, they are all first set to \emph{NaN}.
The bad pixel values are then replaced by the median of the neighbours, ignoring any \emph{NaN} pixels. This ensures that bad pixel
clumps are properly corrected.

\subsection{Point spread function subtraction}\label{sec:psf_sub}

The ``point spread function subtraction'' is the core of the reduction process. It is based on the well established ADI algorithm which
aims at subtracting the stellar point spread function and speckles by using the field rotation in order to
increase the sensitivity to surrounding point sources \citep{marois_angular_2006}.
We are generating a specific point spread function for every single frame, in a similar way to LOCI \citep{lafreniere_new_2007} but
without using any combination coefficients in an effort to preserve the flux.

For an observing sequence composed of $C$ datacubes containing each $N$ frames.
The total number of frames is then $T=C\cdot N$. The frame $f_i=(c_j;n_k)$ will have a parallactic angle
$\alpha_i$ given by
\[
\alpha_i=\arctan\left( \frac{\cos \phi \sin h_i}{\cos\delta\sin\phi - \sin\delta \cos\phi\cos h_i}\right),
\]
with $\phi$ the observatory latitude, $\delta$ the target declination, and $h_i$ the hour angle at observing time $t_i$.

A point spread function is generated for frame $f_i$ using the frames $f_x$ fulfilling the conditions on maximum time separation $t_{\mbox{max}}$
\[
\vert t_k - t_i \vert < t_{\mbox{max}},
\]
and  minimum field rotation $\alpha_{\mbox{min}}$ such that
\[
\vert \alpha_k - \alpha_i \vert > \alpha_{\mbox{min}}=2\cdot\sin\left(\frac{n_{\textsc{\tiny FWHM}}\cdot{\textsc{\tiny FWHM}}}{2\cdot r_{\mbox{min}}}\right),
\]

where $r_{\mbox{min}}$ is the minimum radius in pixels to consider, FWHM is the point spread function full width at half maximum from the
fitting (\ref{sec:registration}),
and $n_{\textsc{\tiny FWHM}}$ is the minimum number of point spread function displacements to prevent companion self-subtraction.

\subsubsection{Fourier shift}

All the geometric operations on the image are based on one and two dimensional Fourier transforms.
As a short reminder, we give the definition of Fourier transforms in two dimensions
of a function $f(x,y)$:
\[\label{eq:2d_fourier}
\hat{f}(\nu_x,\nu_y)=\iint^{\infty}_{-\infty}f(x,y)\cdot{\mbox{e}}^{-i2\pi(\nu_x x + \nu_y y)}dx dy,
\]
and its inverse Fourier transform:
\[\label{eq:2d_inv_fourier}
f(x,y)=\widehat{f}(\nu_x,\nu_y)^{\vee}=\iint^{\infty}_{-\infty}\widehat{f}(\nu_x,\nu_y)\cdot{\mbox{e}}^{i2\pi(\nu_x x + \nu_y y)}d\nu_x d\nu_y.
\]

The one dimensional case is a trivial simplification of the 2D case, and the notation for a one dimensional Fourier transform along the $x$ axis will
be noted by $\hat{f}(\nu_x,y)$, similarly the transform along the $y$ axis will be noted $\hat{f}(x,\nu_y)$.

To perform a shift of the image we use the translation property of Fourier transforms.
If $\widehat{f}(\nu)$ is the Fourier transform of the one dimensional function $f(x)$, then the Fourier transform of
$f(x+a)$ is $\exp(-i2\pi\nu a)\widehat{f}(\nu)$. Thus a spatial shift is equivalent to multiply the Fourier
transform $\widehat{f}(\nu)$ by a phasor ${\mbox{e}}^{-i2\pi\nu a}$. A shift along the $x$ axis in the two dimensional case
can thus be expressed as

\[
f_x(x+ a,y)=
\int^{\infty}_{-\infty}{\mbox{e}}^{-i2\pi\nu_x  a}\widehat{f}(\nu_x,y){\mbox{e}}^{i2\pi(\nu_x x)}d\nu_x, 
\]

and the more general case of a shift $a$ in $x$ and $b$ in $y$ is then obtained by a multiplication by a phasor ${\mbox{e}}^{-i2\pi(\nu_x  a+\nu_y b)}$:

\[\label{eq:2d_shift}
f(x+a,y+b)=\iint^{\infty}_{-\infty}{\mbox{e}}^{-i2\pi(\nu_x  a+\nu_y b)}\widehat{f}(\nu_x,\nu_y)\cdot{\mbox{e}}^{i2\pi(\nu_x x + \nu_y y)}d\nu_x d\nu_y.
\]

The frames we are re-centring are defined on a finite area, whereas the shift property holds for infinite domains. We can
nonetheless apply this operation to the frames provided we introduce a zero-padding which also prevents apparition of
Gibbs oscillations at the borders, but this implies that the operations have to be applied on at least double sized frames, which impacts significantly computational time.

\subsection{Derotation}\label{sec:derotation}
The final step of the ADI processing is to correct each frame for the field rotation and merge all the frames. In order
to preserve the companion signal, and also to keep the noise structure unchanged we perform the rotation using
Fourier transforms.

The rotation algorithm we are using is implemented by applying to Fourier transforms the property that a rotation
matrix can be decomposed in three shear matrices (e.g. \citealt{unser_convolution-based_1995, eddyy_improved_1996, welling_rotation_2006}).
This method is largely used in satellite imagery of the Earth and medical imaging. The galaxy image decomposition tool
GALPHAT is an example of its previous use in astronomical imaging \citep{yoon_new_2011}. We will only give a brief
description of the method based on the detailed
description of the algorithm given by \cite{larkin_fast_1997}. 

Any rotation matrix $R_\theta$ of a given angle $\theta$ can be expressed as the product of three shear matrices:

\[
\underbrace{\begin{pmatrix}
  \cos\theta & -\sin\theta \\
  \sin\theta & \cos\theta
 \end{pmatrix}}_{R_{\theta}}=
\underbrace{\begin{pmatrix}
  1 & -\tan\frac{\theta}{2} \\
  0 & 1
 \end{pmatrix}}_{S_x}
\underbrace{\begin{pmatrix}
  1 & 0 \\
  \sin\theta & 1
 \end{pmatrix}}_{S_y}
\underbrace{\begin{pmatrix}
  1 & -\tan\frac{\theta}{2} \\
  0 & 1
 \end{pmatrix}}_{S_x}
\]

where $S_x$ and $S_y$ are shear matrices on the $x$ axis and $y$ axis respectively.

To shear by a factor $a=\tan\frac{\theta}{2}$ in the $x$
direction an image described by the function $f(x,y)$ we apply the transformation $s_x(x,y)=f(x+a y,y)$, which can be
readily adapted to Fourier transforms using their shift property. The shear matrix $S_x$ applied to the image $f(x,y)$
can then be expressed in terms of Fourier transforms as the function $s_x$

\[
s_x(x,y)=
\int^{\infty}_{-\infty}\overbrace{\underbrace{{\mbox{e}}^{-i2\pi\nu_x a y}\widehat{f}(\nu_x,y)}_{\text{FT}}\cdot{\mbox{e}}^{i2\pi\nu_x x}d\nu_x}^{\text{IFT}}\ , 
\]

the product $S_y S_x$ becomes by noting $b=-\sin\theta$
\[
s_{yx}(x,y)=
\int^{\infty}_{-\infty}{\mbox{e}}^{-i2\pi\nu_y b  x}\widehat{s}_x(x,\nu_y)\cdot{\mbox{e}}^{i2\pi\nu_y y}d\nu_y, 
\]

and the rotation $S_x S_y S_x$
\[
s_{xyx}(x,y)=
\int^{\infty}_{-\infty}{\mbox{e}}^{-i2\pi\nu_x a y}\widehat{s}_{yx}(\nu_x,y)\cdot{\mbox{e}}^{i2\pi\nu_x x}d\nu_x .
\]

Similarly to the shift case, the frames need to be padded. This implies that in order to rotate one frame 6 FFT have to be
applied on a double sized frame, resulting in a significant increase of computation time compared to standard interpolation methods.
This rotation technique can only be applied to such a large amount of frames thanks to parallelisation.

\section{Pipeline performance}

\begin{figure}
\centering
\includegraphics[width=0.5\textwidth]{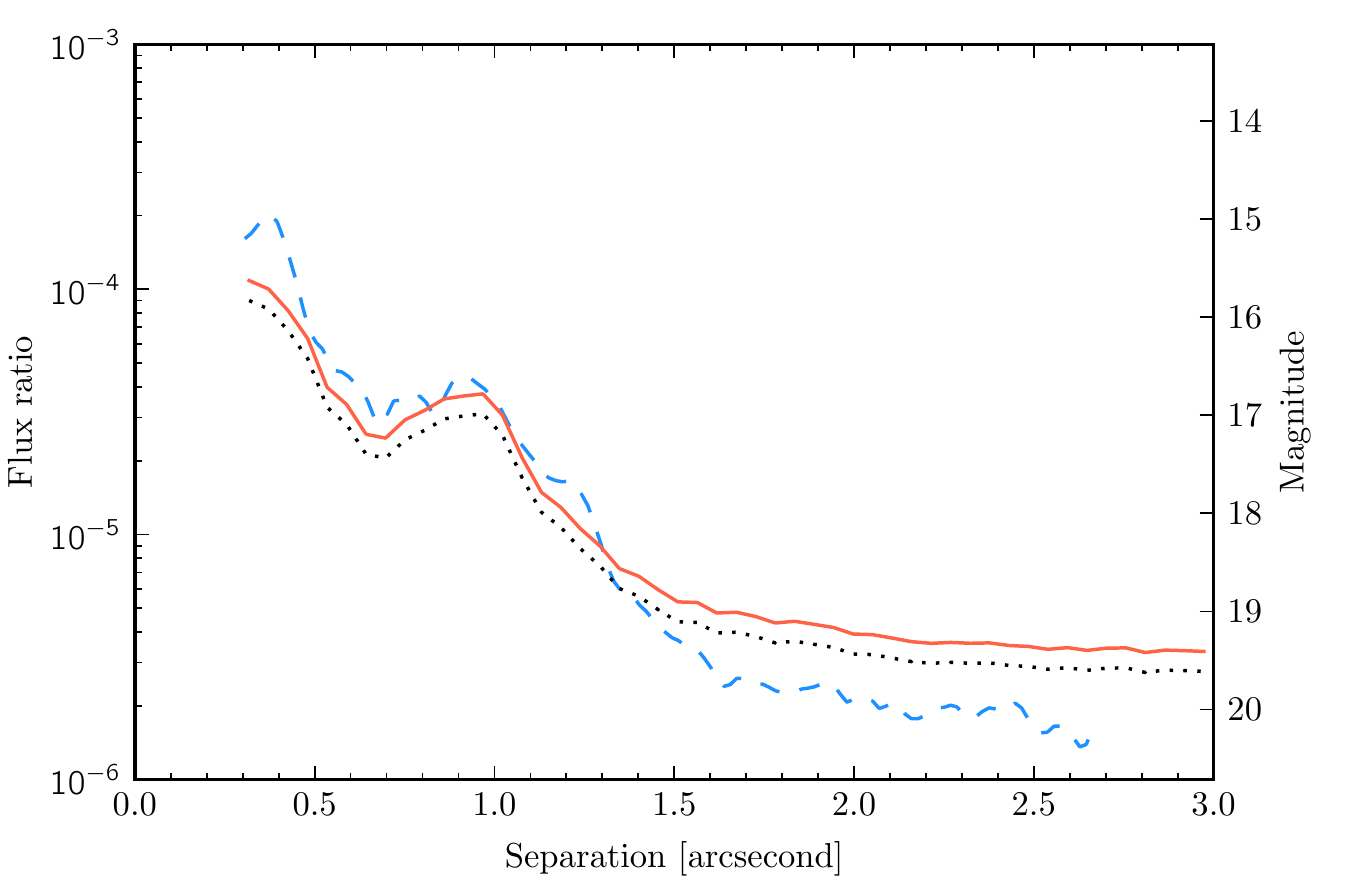}
\caption{Angular differential imaging detection limits for Gemini/NICI data in \textit{CH4-K5\%S} band, using a $0.22 \arcsec$
  semi-transparent coronographic mask
  with $\approx\,40\degr$ of field rotation. The three lines are detection limits at 5-$\sigma$.
  The black dotted line is the detection limit achieved by {\small GRAPHIC} before correction
  for self-subtraction while the red solid line is the detection limit of {\small GRAPHIC} corrected in order to take into account the flux loss.
  The blue dashed line is the detection limit using a principal component analysis pipeline which was used in \citet{casassus_flows_2013}.
\label{fig:nici_detlim}}
\end{figure}

Contrast curves obtained for HD142527 observations with NICI, using {\small GRAPHIC} and PCA are displayed in \autoref{fig:nici_detlim}. At small separation we reach a higher contrast than PCA, while
the low-pass filtering from interpolation results in better contrasts at higher separation where the noise is mainly Gaussian.

To test the pipeline performance we developed an algorithm to inject fake companions. These companions are generated by
first including their signal into a plane wave-front which is then convoluted with a pupil based on the main optical
characteristic of the used telescope. Using this technique we have a precise control on the companion flux, and furthermore the point spread function
scales precisely with wavelength. Poisson noise is finally added to these nearly perfect point spread functions before adding them to the
real science frame.

\subsection{Performance of geometric transformations}

To characterise the performance of the shift and rotation algorithms, we generated a test image with fake companions.
This image is composed of an L' short-exposure image with a saturated point spread function, which has been sky-subtracted, cleaned from
bad pixels, and median-filtered.
To this image we added fake companions with magnitudes differences to the star reaching step-wise from 1 to 8 and with
separations from 0.5 to 6.5 arcseconds with arcseconds steps, and Poisson noise was included in the process.

\begin{figure}
\centering
\includegraphics[width=0.5\textwidth]{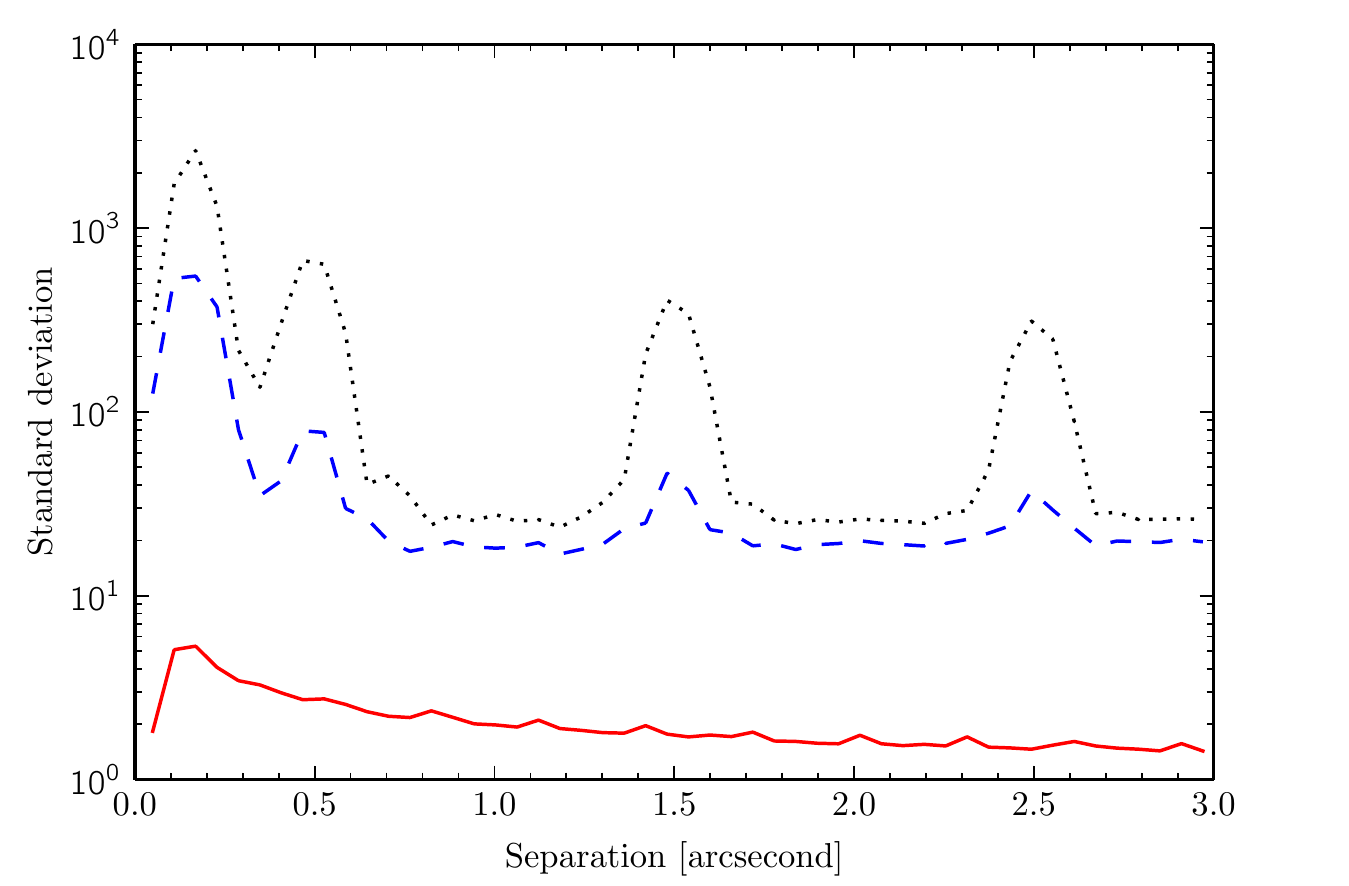}
\caption{Root mean square representing the noise caused by the shift algorithms. As a reference the root mean square of the
  original image as a function of separation to the centre is given by the black dotted line. To test the algorithms we shift an image
  first by $\Delta x=3.5, \Delta y=2.7$ pixels and then back. The difference between the original image and the one shifted back to the
  initial position is then plotted for the interpolation and Fourier shift, blue dashed and red solid lines respectively.
\label{fig:2shift_test_detlim}}
\end{figure}

To test the performance of the shift algorithm, we shifted the original image in a $\Delta x, \Delta y$ direction and
then shifted it back in the opposite direction ($-\Delta x, -\Delta y$). This double shifted image can then be compared with the
original non-shifted image. By subtracting the original image from the double-shifted, the
effects induced by the different shifting algorithms become visible. The dotted black line in
\autoref{fig:2shift_test_detlim} shows the normalised root mean square of the original test frame, calculated in concentric
annuli. The injected fake companions are causing the peaks at 0.5, 1.5, and 2.5 arcseconds. The two additional lines
show the root mean square of the difference between the original test frame and the double shifted frames, where the interpolation
and Fourier shift are represented by the blue dashed and red solid lines respectively. These two lines would
be flat with no root mean square if the shift algorithms were perfect. For the spline interpolation
this is clearly not the case, with the curve reflecting two phenomenons in and out of the peaks which are both caused by
the fact that interpolations in image space act as an uncontrolled low-pass filter. When the image is interpolated, the
structure of the noise is modified so that the high frequency noise is not removed by the subtraction, as it is missing
from the double shifted frame. This effect leaves an overall noise
continuum. The effect on the peaks comes from the fact that the fake companions are flattened by the interpolation, so that
part of the fake companion signal is not removed.

\begin{figure}
\centering
\includegraphics[width=0.5\textwidth]{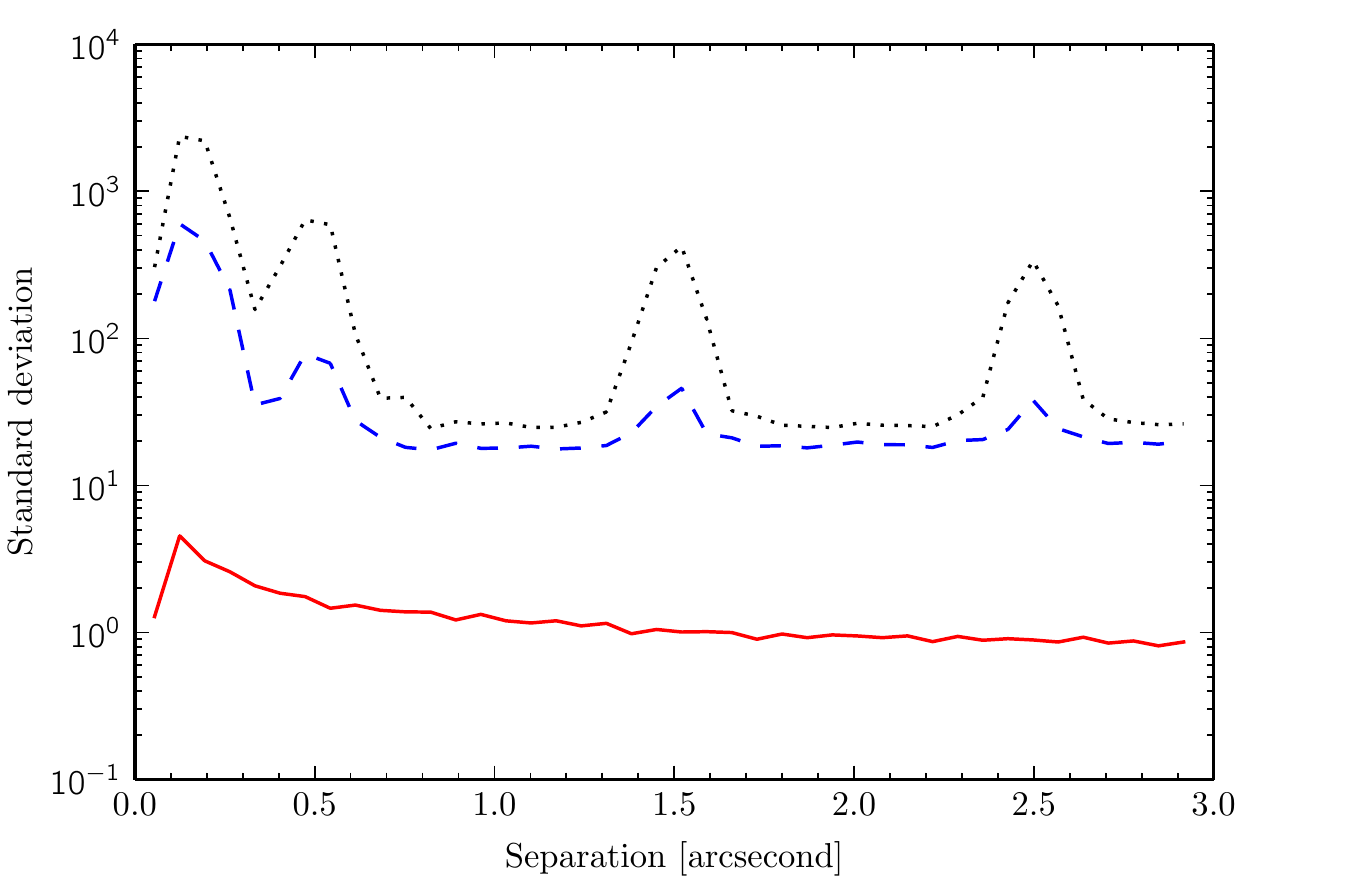}
\caption{
  Root mean square representing the noise caused by the rotation algorithms. As a reference the root mean square of the
  original image as a function of separation to the centre is given by the black dotted line. To test the algorithms we rotate the image
  first by an angle $\alpha=11.3$ and then back. The difference between the original image and the one rotated back to the
  initial position is then plotted for the interpolation and Fourier rotation, blue dashed and red solid lines respectively.
\label{fig:2rot_test_detlim}}
\end{figure}

For the rotation algorithm we applied a similar test. We first rotated a test image by an angle $\alpha=11.3$ followed by a rotation
by an angle $-\alpha$. Ideally such a double rotation should return the original image. Changes in the image structure
induced by the rotation can be found by subtracting the original image from the double rotated one. The root mean square  as a function of separation to
the central point spread function of the
original image is plotted on \autoref{fig:2rot_test_detlim} with a black dotted line, and the subtracted rotations using interpolation
and Fourier shears are represented on the same figure by the blue dashed and red solid lines respectively.

\begin{figure*}
\centering
\includegraphics[width=1\textwidth]{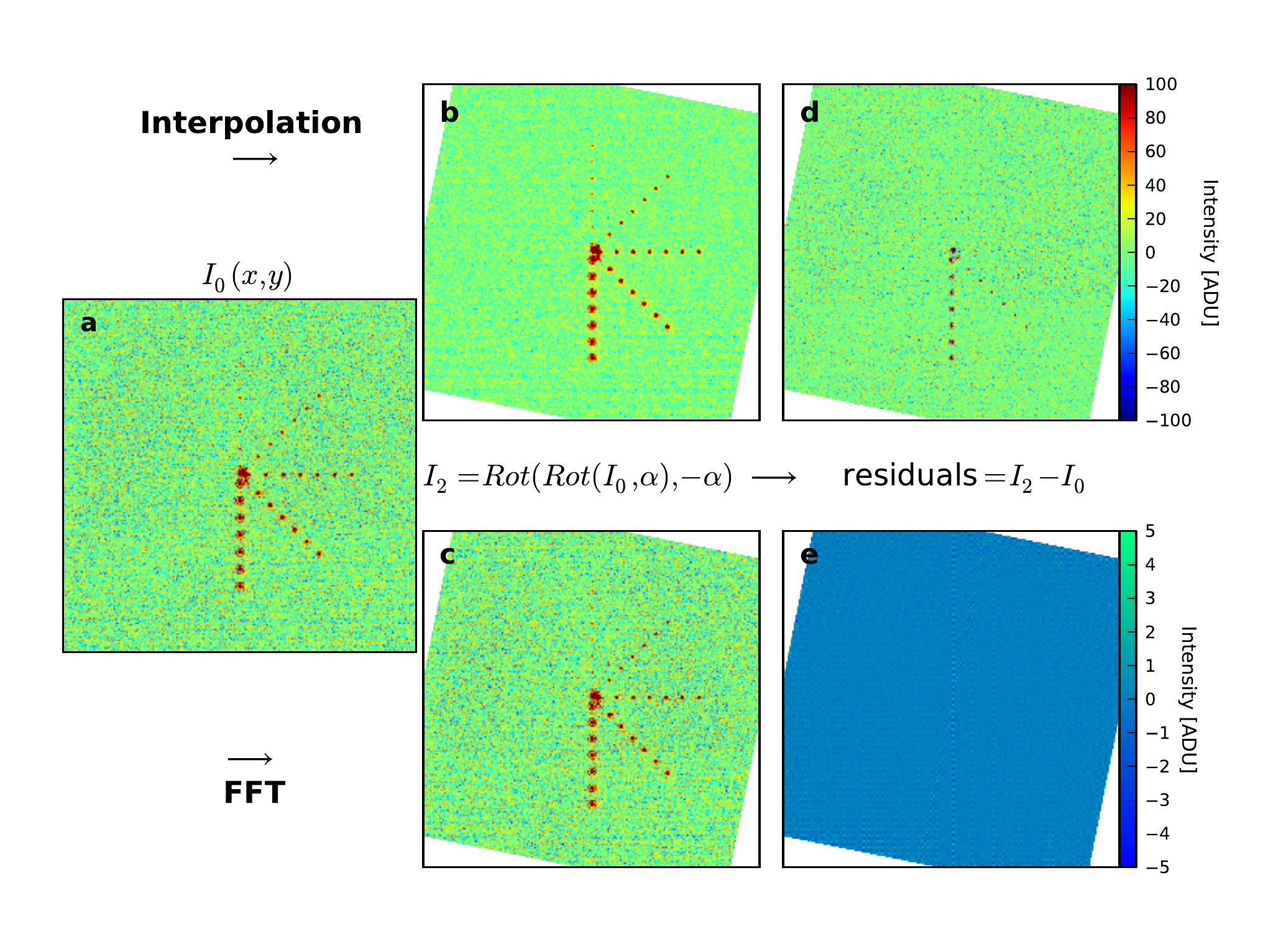}
\caption{The original non-rotated test image is given in figure (a).
The result of two consecutive $11.3\degr$ and $-11.3\degr$ rotations of the original figure (a), using a third order spline interpolation, and the 3 shear algorithm are shown in figure (b) and (c) respectively. The residuals are computed by subtracting the original image from the
interpolation rotated (d), and three-shear rotated (e) images. All panels have the same intensity scale, except for figure (e)
where we reduced the cuts by a factor 20 to reveal some of the induced noise.
\label{fig:10rot_test}}
\end{figure*}

The images from the rotation test are represented on \autoref{fig:10rot_test}, where \ref{fig:10rot_test}a is the
original image, \ref{fig:10rot_test}b and \ref{fig:10rot_test}c are the resulting image from the rotation followed by an
inverse rotation using interpolation and the 3-shear algorithm respectively. The most striking difference between the
two rotated images is the residual noise structure, the 3-shear algorithm preserving the noise structure while the
interpolation algorithm acts as an uncharacterised low-pass filter.

The different effects of the two rotation algorithms is even more evident when the original image is subtracted from the rotated images, as
can be seen on \autoref{fig:10rot_test}d for the interpolation and
\autoref{fig:10rot_test}e for the 3-shear algorithm. In the case of interpolation, the companions become dark blue
points surrounded by signal which means that the companion point spread function is spread, with signal being transferred from the centre of the point spread function
into the wings. The 3-shear residuals show no structure, indicating that the star and companion signals are not altered
by the rotation. Some numerical noise can be noted, but due to its very high frequency it can be filtered out without affecting the companion signal.

\subsection{Photometric accuracy}
Test data sets are created by injecting fake companions into all the raw frames with an angle following the field
rotation. By injecting the companions into the raw frames we are able to take into account nearly all the steps of the
reduction, namely image pre-processing, re-centring, point spread function subtracting, de-rotating, and final collapse. Centre and parallactic
angle determination are the only two operations we cannot test with this method, as the companion injection already
relies on these two parameters.

To illustrate the photometric accuracy we generated such a dataset using a 2 hour \textit{L'} NACO observation sequence with a
$120\degr$ field rotation, centred on the transit through meridian. The \textit{no binning} line in \autoref{tab:bin_flux} shows that in the
worst case we have a flux loss of 40 per cent, which is similar to what is achieved with the PCA pipeline {\small PYNPOINT} but better by a factor 2 compared to {\small LOCI} \citep{amara_pynpoint:_2012}.

Our ability to inject fake companions in the raw frames is thus a key component to recover accurate photometry. By
injecting the companions at the very beginning we can take into account all the effects that could decrease
the companion flux during the various data reduction steps.

\subsection{Effect of binning}

The data set used to quantify the photometric accuracy was also used to analyse the effect of binning. To study this effect
we binned the initial data set by combining each set of ten 0.2 second frames into a single median frame
resulting in a smaller data set. We did the same by also combining 50 frames into a single one representing 10 second total integrations per frame.

\begin{figure}
\centering
\includegraphics[width=0.5\textwidth]{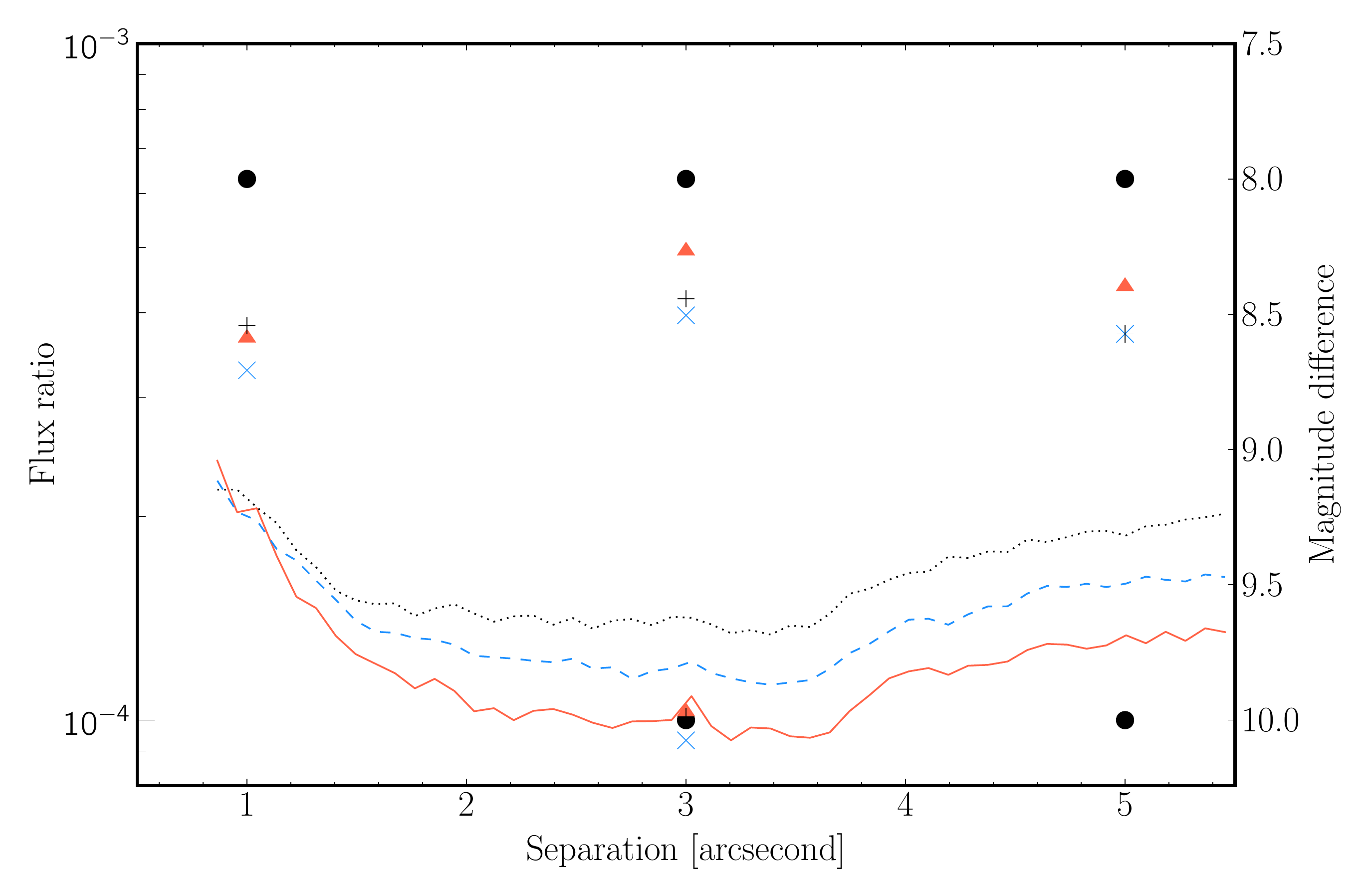}
\caption{Angular differential imaging detection limits for the same initial 2 hour observation sequence ($120\degr$ rotation). The injected fake
companions are represented by the black circles.
The non binned detection
limit and recovered companion flux are given by the red solid line and triangles respectively. The 10 frame bins are represented by the blue dashed
line and triangles, while the 50 bin results are given by the black dotted line and plus signs.
\label{fig:detlim_bin}}
\end{figure}

The three data sets were then reduced using exactly the same parameters. The resulting detection limits, along with retrieved photometry of the fake
companions are plotted in \autoref{fig:detlim_bin}. The detection limits increases beyond $3.5\arcsec$
because the result of dithering is that less frames are available at these separations which decreases the signal to noise ratio.

\begin{table}
\centering
\caption{Magnitude difference of the injected companions and percentage of the recovered flux as a function of separation and bin size.\label{tab:bin_flux}}
\begin{tabular}[t]{l  c r@{ -- } l  r@{ -- } l}
          & $1\arcsec$ & \multicolumn{2}{c}{$3\arcsec$} & \multicolumn{2}{c}{$5\arcsec$} \\\cline{2-6}
Injected $\Delta$mag & 8 & 8 & 10 & 8 & 10\\
Recovered no binning & 60\% & 81\% & 106\% & 71\% & 72\%\\
Recovered 10 frame bins &  52\% & 63\% &93\%  & 59\%& 20\%\\
Recovered 50 frame bins &  61\% & 66\% & 101\% & 60\% &52\%\\
\end{tabular}
\end{table}

The percentage of recovered flux of the fake companions is given in \autoref{tab:bin_flux}. Flux recoveries above 100\% are caused by low signal to noise,
for companions at or below the detection limit.

\subsection{Performance versus observation duration}

In order to test our observing strategy of long observations, we defined a specific test case. Taking one of our
nearly three hour
observation data set, we added fake companions to the raw images. The observation was then reduced using three different data set
sub-samples. For the first one we used the whole data set trimmed in order to have as much observing time before and after
transit at meridian. This results in a two hour data set, with one hour before and one hour after meridian transit.
We did the same for a one hour sub-sample and a 30 minutes sub-sample from the same initial data set, each time centred
on the meridian transit. The selected data set is of a star at a declination of $-17\degr$, which results in a smooth rotation rate.
Targets which are crossing the meridian at zenith will have the entire field rotation happening in only a few minutes.
For such specific targets the gain in sensitivity at small separation for longer observations will be very limited.

\begin{figure}
\centering
\includegraphics[width=0.5\textwidth]{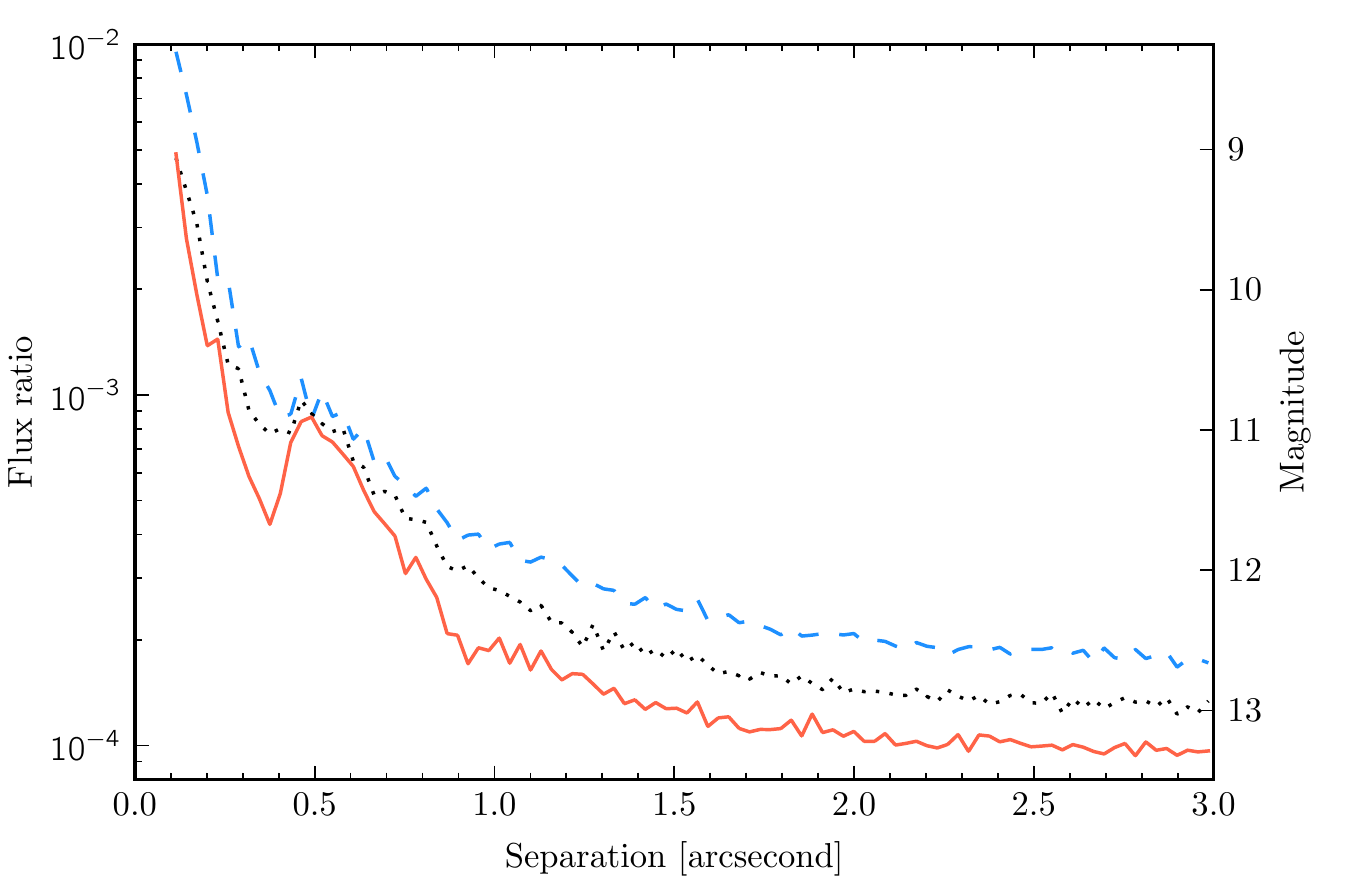}
\caption{Angular differential imaging detection limits for an observation in \textit{L'} band of a 6.5 magnitude star at $-17\degr$ declination. The data sets taken from this sequence are all centred on
meridian transit using 30 minutes ($\approx 50^\circ$ rotation), one hour
($\approx 90^\circ$) and two hours sub-samples ($\approx 140^\circ$), plotted in blue dashed, black dotted, and solid red line respectively.
\label{fig:detlim_3times}}
\end{figure}

The resulting detection limits obtained by reducing the three sub-samples with exactly the same reduction parameters are plotted
in \autoref{fig:detlim_3times}. The detection limits clearly show that the reduction was tuned for the inner region within
0.5 arcseconds. At 0.3 arcseconds separation, the achieved magnitude difference are 7, 7.6, and 8 for 30 minutes, one hour,
and two hours respectively. At 1.5 arcseconds separation these limits become 9, 9.3, and 9.7. With two hour observations we
thus gain one magnitude in sensitivity with respect to a short 30 minutes sequence, and half a magnitude with respect to
a conventional one hour observation. A one magnitude difference in \textit{L'} band is what separates a 13 $M_J$ from a 20 $M_J$ companion,
based on \citet{allard_bt-settl_2014} at 1 Gyr.

\section{Conclusions}

We have presented a new pipeline for angular differential imaging which is the first, to our knowledge, to perform the critical steps of image re-centring and field de-rotation in the Fourier domain and on individual frame. By doing so, we reduce most of the image smoothing introduced by interpolation and frame binning. The resulting image noise characteristic is also preserved. This theoretical approach is validated by several test cases, where we show that excellent dynamic range and  photometry retrieval are obtained at separations shorter than 1.5 arcsec compared to a standard PCA analysis.

In addition, we validated the observing strategy that consist in observing the same target for at least two consecutive hours with a clear boost in performances of 1 magnitude at short separations in the case of a target with a $120\degr$ rotation in 2 hours, representative of a typical ADI target. {\small GRAPHIC} is coded in python with a few C-modules and was designed to run on HPC clusters, with a minimum requirement of 2 cores and  4GB RAM per core. Provided the right environnement (48 cores, 192GB RAM), {\small GRAPHIC} is able to process up to 100'000 2048x2048 frames with no binning thanks to massive parallelism in 9 hours. This parallelisation of the pipeline makes it also possible to implement further computationally demanding algorithms. Further development is planned in order to use graphics processing units (GPU) for a gain in processing time and wavelet filtering. {\small GRAPHIC} has also very recently been adapted to process VLT/SPHERE and Subaru/SCExAO data. {\small GRAPHIC} has also been applied on an extended source with results published in \citet{casassus_flows_2013} under the development name {\small PADIP}.

\section*{Acknowledgements}
The authors would like to thank J. Carson for his refereeing work and for the
many useful comments he has given, and
D. Mawet for kindly providing PCA detection limits from \citet{casassus_flows_2013}.
J.H. is supported by the Swiss National Science Foundation (SNSF).
To develop the pipeline the author also made use of {\small SCIPY} \citep{jones_scipy:_2001}, {\small  NUMPY} \citep{oliphant_python_2007},
 {\small ASTROPY}
\citep{astropy_collaboration_astropy:_2013},  {\small BOTTLENECK},  {\small IPYTHON} \citep{perez_ipython:_2007}, and
 {\small MATPLOTLIB} \citep{hunter_matplotlib:_2007}.

\bibliographystyle{mn2e_fix}

\inputencoding{latin1}
\bibliography{reduction_pipeline}
\inputencoding{utf8}

\bsp

\label{lastpage}
\end{document}